\documentclass[a4paper]{jpconf}
\usepackage{graphicx}
\usepackage{url} 

\newcommand{\arcdeg}{\mbox{$^\circ$}}

\begin{document}
\title{An African VLBI network of radio telescopes}

\author{M. J. Gaylard$^1$, M. F. Bietenholz$^{1,2}$, L. Combrinck$^1$, R. S. Booth$^{3,4}$, S. J. Buchner$^1$,
B. L. Fanaroff$\,^3$, G. C. MacLeod$^5$, G. D. Nicolson$^1$, J. F. H. Quick$^1$, P. Stronkhorst$^1$, T. L.
Venkatasubramani$^3$}

\address{$^1$ HartRAO, P. O. Box 443, Krugersdorp 1740, South Africa}
\address{$^2$ Department of Physics and Astronomy, York University, Toronto,
Ontario, M3J 1P3, Canada}
\address{$^3$ SKA South Africa, 17 Baker St., Rosebank, Johannesburg, South Africa}
\address{$^4$ Onsala Space Observatory, SE-439 92 Onsala, Sweden}
\address{$^5$ Department of Science and Technology, Private Bag X894, Pretoria 0001, South Africa}

\ead{mike@hartrao.ac.za}

\begin{abstract}
The advent of international wideband communication by optical fibre has
produced a revolution in communications and the use of the internet. Many African
countries are now connected to undersea fibre linking them to other African countries
and to other continents. Previously international communication was by microwave
links through geostationary satellites. These are becoming redundant in some
countries as optical fibre takes over, as this provides 1000 times the bandwidth of the
satellite links.

In the 1970's and 1980's some two dozen large (30~m diameter class) antennas
were built in various African countries to provide the satellite links. 
Twenty six are currently known in 19 countries. As these
antennas become redundant, the possibility exists to convert them for radio
astronomy at a cost of roughly one tenth that of a new antenna of similar
size. 

HartRAO, SKA Africa and the South African Department of Science
and Technology (DST) have started exploring this possibility with some
of the African countries.
\end{abstract}

\section{Introduction}

New undersea fibre optic cables provide communication bandwidths a thousand
times greater than radio links via communications satellites.  Large
antennas at satellite Earth stations are becoming redundant as a result. 
This trend has now reached Africa with the advent of fibre optic cables down
the East and West coasts of the continent.

More than two dozen large Satellite Earth Station antennas built in the
1970's and 80's exist in Africa.  Twenty six are currently known in 19 countries. 
Antennas that are redundant could be
refitted for radio astronomy at comparatively small cost.

Africa currently has only one antenna equipped to participate in global
radio astronomy using Very Long Baseline Interferometry (VLBI) for high
resolution imaging, this being the 26~m ex-NASA antenna at Hartebeesthoek in
South Africa. The window of opportunity to create an African network of
VLBI-capable radio telescopes from redundant large satellite antennas exists
now. This African VLBI Network (AVN or AfVN) would initially work with existing VLBI
arrays such as the European VLBI Network (EVN), but could operate
independently if sufficient antennas become available.

Such a network would bring new science opportunities to participating
countries on a short time scale, enable participation in SKA pathfinder
development and science, and would help create the environment for bringing
the Square Kilometre Array (SKA) to Africa.  It would provide a ready made
VLBI extension for the MeerKAT radio telescope currently being developed as an
SKA precursor.

\section{Intelsat Satellite Earth Stations}

\subsection{Intelsat Standard A Satellite Earth Station antennas}

From the 1960's, communication via satellites orbiting the Earth was
introduced to carry voice, data and TV signals, to supplement undersea
cables.  The radio bands allocated for this were 5.925 $-$ 6.425 GHz
for uplink and 3.700 $-$ 4.200 GHz for downlink.  These are within the
frequency range known as ``C-band''. Intelsat defines Earth Stations
for this band as Standard A, B, F or H depending on their technical
characteristics \cite{Intelsat IESS-207}.  Initially a Standard A
antenna had to be at least 30~m in diameter, and the antennas built in
Africa from 1970 to 1985 are this size.  From 1985 new technology
enabled satellite transmitter power to increase and the required size
was greatly reduced.

\subsection{Large Satellite Earth Station antennas in Africa}

\begin{figure}
\begin{center}
\includegraphics[width=25pc]{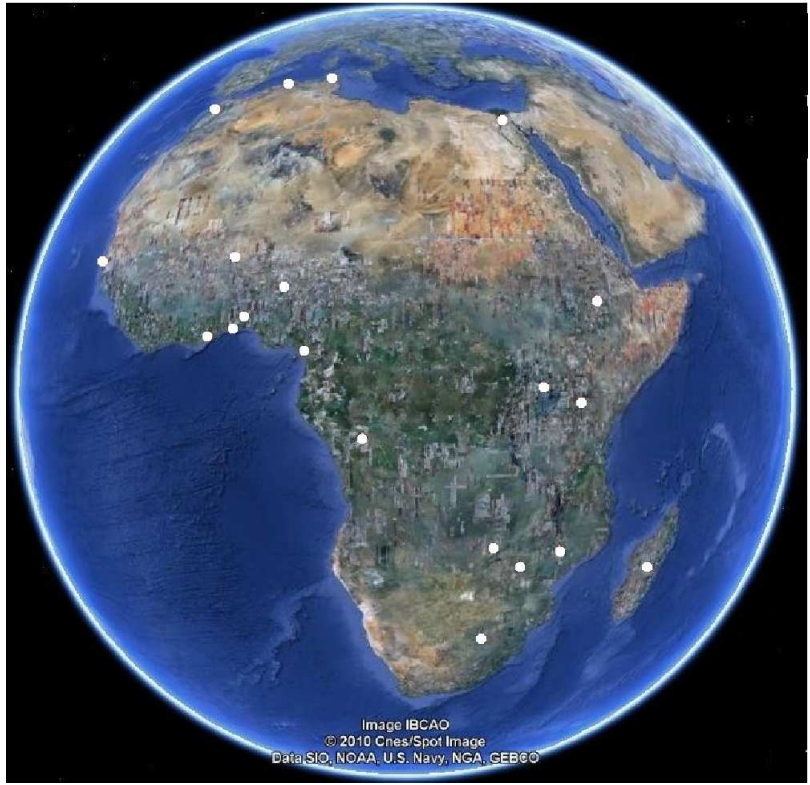}
\end{center}
\caption{\label{fig:antenna_map}Locations of identified large Satellite
Earth Station antennas in Africa}
\end{figure}

SKA Africa partner countries identified with large antennas are South
Africa (3 antennas), Ghana, Kenya, Madagascar and Zambia.  Other
African countries in which large antennas have been located are
Algeria (2), Benin, Cameroon (2), Congo Peoples Republic, Egypt (2),
Ethiopia, Malawi, Morocco, Niger, Nigeria (3), Senegal, Tunisia,
Uganda and Zimbabwe.  These two groups are shown in
Fig.~\ref{fig:antenna_map}.  In addition, Gabon probably has one and
one is reported in the Congo Democratic Republic.  The antenna in
Mozambique was dismantled and probably also the one in Togo.  South
Africa's SKA partners in Africa are Botswana, Ghana, Kenya,
Madagascar, Mauritius, Mozambique, Namibia and Zambia.

\subsection{Large Satellite Earth Station antennas outside Africa converted for radio
astronomy}

Operational converted antennas are the 30~m Ceduna antenna in Australia
\cite{ceduna} and the 32~m Yamaguchi antenna in Japan \cite{yamaguchi}.  The
Warkworth 30~m antenna in New Zealand was handed over for conversion on
19 November 2010 \cite{warkworth}.  In Peru the Sicaya 32~m antenna is being converted
with assistance from the Yamaguchi team and saw first light in March 2011
\cite{peru}.  Three antennas at the Goonhilly Downs station in the UK
decommissioned in 2008 are proposed for conversion for use with e-Merlin and
the EVN \cite{goonhilly}.  On 10 May 2011 it was announced that the 32~m
Elfordstown antenna outside Cork in Ireland is to be converted
\cite{ireland}.

\section{Astronomical applications for redundant large antennas in Africa}

\subsection{Astronomical Very Long Baseline Interferometry (VLBI)}

Widely separated radio telescopes operating together create a virtual
telescope equal in size to the project distance, or baseline, between the
telescopes.  The angular resolution of a telescope depends on its size, so
the larger the separation, the better the resolution.  Practically, to
create good images a number of telescopes are needed, separated by small,
medium and large distances.

The HartRAO 26~m radio telescope is valuable in providing long
baselines to radio telescope arrays in on other continents,
e.g. Europe (European VLBI Network - EVN) and Australia (Australia
Telescope Long Baseline Array - AT-LBA), and thus high angular
resolution imaging.  However, the large distance between the South
African telescope and the others makes for less than ideal imaging.
It would be helpful to have antennas filling the gap between South
Africa and Europe and South Africa and Australia (for example) and the
image quality would be substantially improved.  They would be a
powerful addition to the radio telescope arrays on other continents.

To create an interferometer able to operate independently to produce images,
a minimum of four antennas are needed to provide both phase and amplitude
closure.  Thus a minimal African VLBI Network able to operate independently
of the radio telescopes on other continents could be formed if four suitable
antennas were  brought into use on the continent and on neighbouring
islands.  Science capability would improve substantially if more large
antennas could join the network.  With four antennas, 50\% of the
phase information and 33\% of the amplitude information can be recovered;
with ten antennas these rise to 80\% and 78\% respectively \cite{Pearson}.

Astronomical objects suitable for study by VLBI are those that are
radio-bright and of small angular size.  Objects of very large physical size
meet this requirement if they are quite distant, and radio galaxies and
quasars are examples.  Radio-bright supernovae in external galaxies are good
examples of objects whose evolution can be studied with VLBI.  Masers in
star-forming regions in the Milky Way are examples of nearby bright, compact
sources.  Methanol masers at 6.668 GHz and 12.178 GHz are of particular
interest currently.  Repeated mapping would help in investigating the causes 
of their variability, especially those found to be showing regular variations. 
Measurement of their annual parallaxes by repeated VLBI
observations enables their distances to be determined, and thus the
locations of the spiral arms in the Milky Way, where these occur, to be
measured accurately.  This is only being done by northern VLBI arrays
currently, leaving the fourth quadrant of the Milky Way, in the far southern
skies, so far unmapped by this method.

Distances to pulsars can be obtained by astrometric VLBI.  Transient sources
and gamma-ray bursts are all potential targets.  The on-going study of
southern hemisphere radio galaxies (TANAMI project) would be enhanced with
extra telescopes able to see the southern skies.  Microquasars in the Milky
Way are targets, as are radio-loud interacting binary stars such as Circinus
X-1.

Spacecraft in interplanetary space and at other planets that are equipped
with 8.4 GHz frequency-stable transmitters are being used for precise
spacecraft position determination and for study of the interplanetary
medium.  This technique is known as the Planetary Radio Interferometry and
Doppler Experiment (PRIDE).  The HartRAO 26~m telescope, and others, are
regularly observing the Venus Express (VEX) spacecraft in orbit around Venus
for this purpose.   The Russian Phobos-Grunt mission to Mars' moon Phobos is
due for launch late in 2011 and will use this technique.  Converted
satellite antennas with 8.4 GHz receivers in Africa would be good for this
owing to their collecting area and location near the equator.

Antennas equipped with dual 2.3+8.4 GHz receivers would be able to participate 
in geodetic and astrometric VLBI with the current generation of radio telescopes 
carrying out this research.  This technique provides a very precise absolute 
location for the telescope, which can be transferred to co-located relative 
position measuring systems such as the global navigation satellite systems GPS 
and GLONASS, and the laser ranging and DORIS systems used to obtain precise
measurements of satellite orbits.

\subsection{Single-dish astronomy}

The only VLBI array that runs essentially continuously is the US Very Long
Baseline Array (VLBA).  The EVN now runs much more frequently than the
traditional three four-week sessions per year of the past, with monthly e-VLBI
(real-time data streaming to the correlator) and
Target Of Opportunity (TOO) VLBI.  However there would still be substantial
non-VLBI time available for operation of each telescope as a stand-alone instrument.  

The time available for single-dish astronomy would be valuable for student 
training purposes and for selected
research projects.  Techniques that could be developed would be radiometry -
measuring the brightness of broad-band emission sources, spectroscopy -
measuring emission and absorption line strengths, and pulsar timing -
measuring the arrival times of pulses from neutron stars.  Establishing a
capability for unattended queue-scheduled single-dish observing (as on the
HartRAO 26~m telescope) would greatly assist in permitting time series to be
built up of the behaviour of scientifically interesting variable sources
that can be observed with these techniques.

\section{Towards the African VLBI Network}

\subsection{Identifying candidate antennas}

Intelsat documentation and internet mapping services using high resolution
photography of the Earth's surface enabled the current existence of the
large antennas to be confirmed.  Postage stamps commemorating their
inauguration were another valuable source of information.  Government level
contact has been made with some of the countries by DST to further investigate the
possibility of conversion.  Ghana was one of those approached, and was quick
to seize the opportunity.  This opportunity is a direct result of the African SKA Project.

\subsection{The Kuntunse Satellite Earth Station in Ghana}

\begin{figure}[h]
\begin{minipage}{18pc}
\includegraphics[width=18pc]{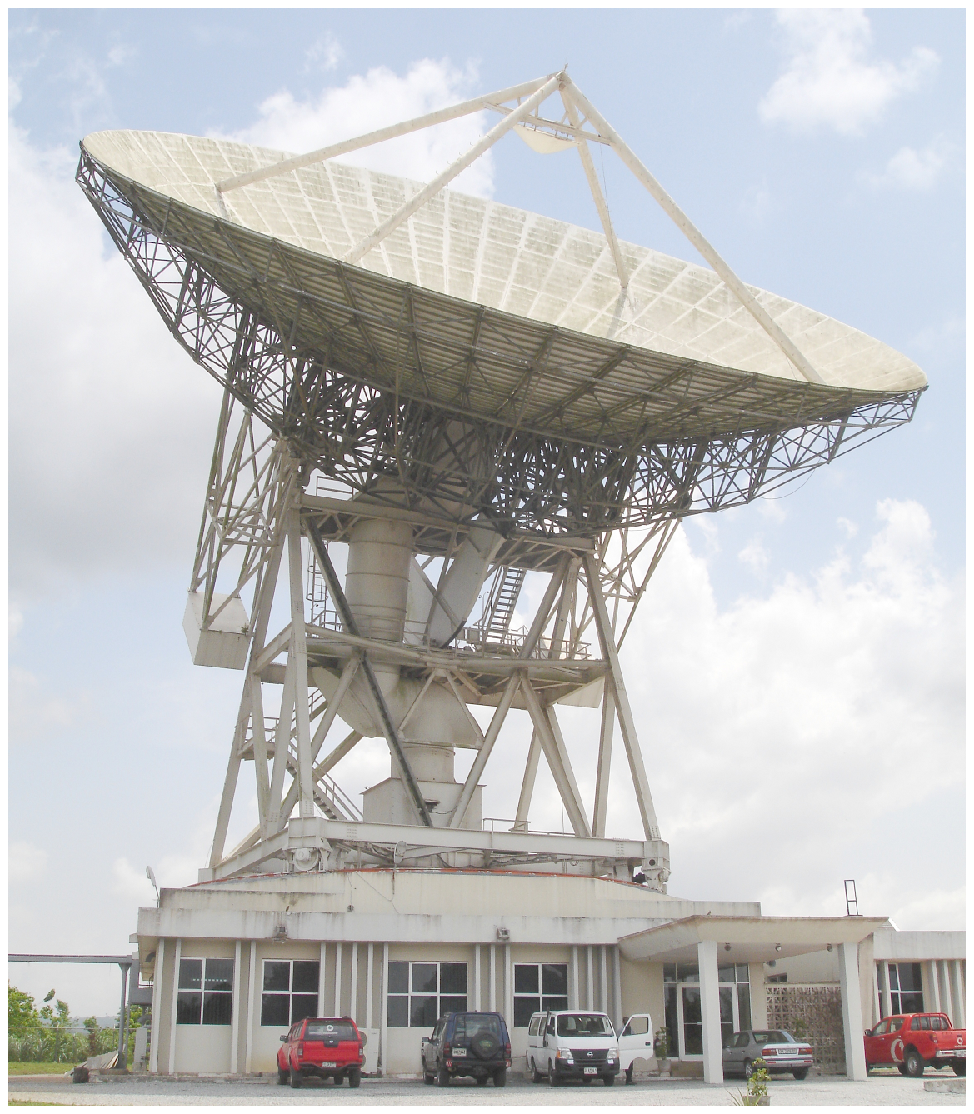}
\caption{\label{fig:kuntunse}The Kuntunse 32~m antenna photographed in March
2011.}
\end{minipage}\hspace{1.8pc}%
\begin{minipage}{18pc}
\includegraphics[width=18pc]{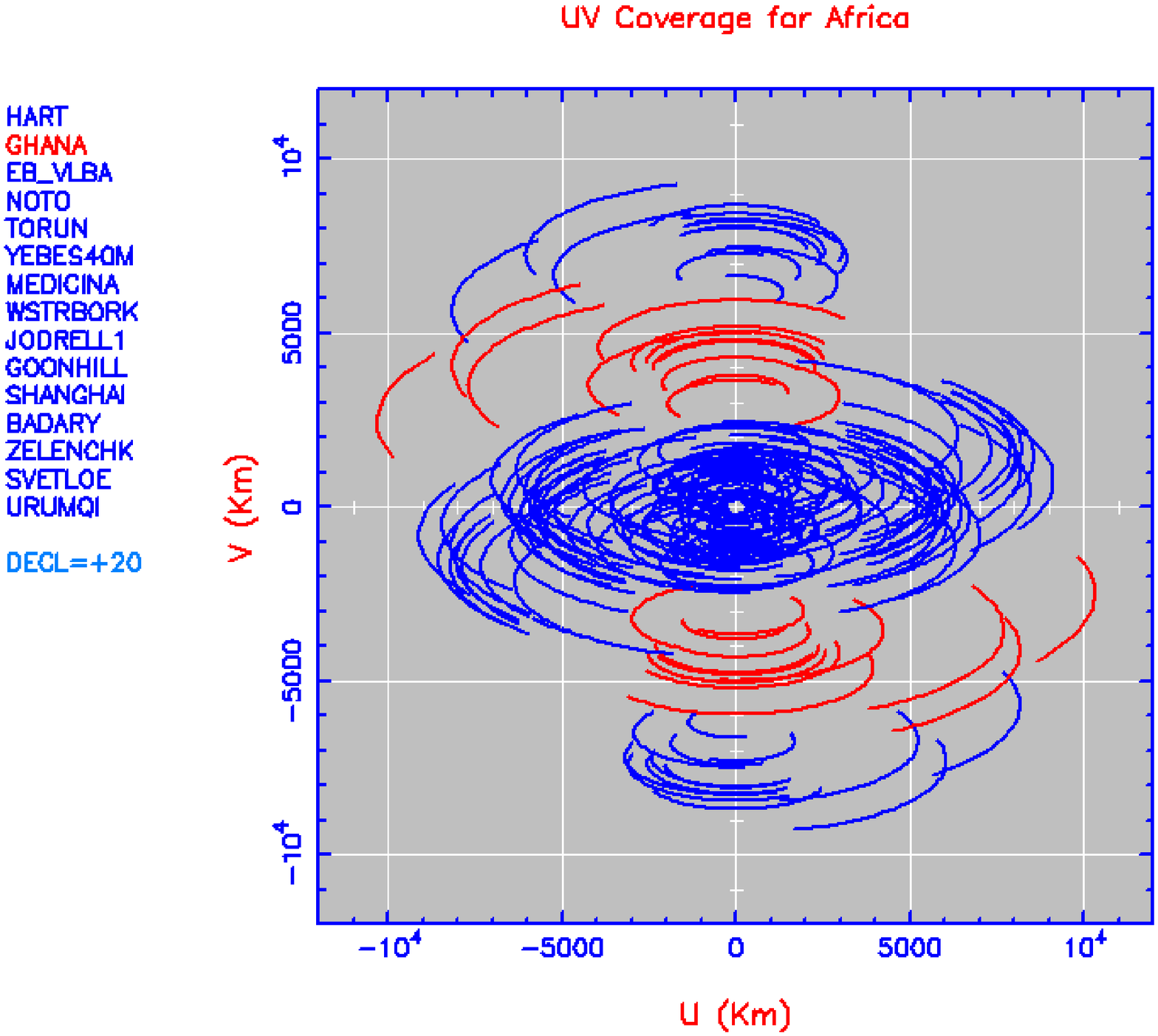}
\caption{\label{fig:UV_kuntunse}VLBI UV diagram for a source at +20\arcdeg\
declination. The red tracks show how the Kuntunse antenna
improves the UV coverage by filling the gap between Europe and South Africa.}
\end{minipage} 
\end{figure}

The Ghanaian government and Vodafone, the owners of the Kuntunse Satellite
Earth Station outside Accra, have agreed to convert the 32~m antenna at the
station for radio astronomy.  This antenna has been out of service for about
ten years; an adjacent 16~m antenna is carrying the remaining satellite
traffic.  A HartRAO team and Prof.\ Michael Jones from Oxford University
visited Kuntunse in March 2011 for an initial feasibility investigation,
which indicated that it was in a suitable condition to proceed
(Fig.~\ref{fig:kuntunse}).  A HartRAO/SKA Africa team visited in May 
to assist in starting the rehabilitation.

This antenna is located roughly half way between South Africa and Europe.
Fig.~\ref{fig:UV_kuntunse} shows how it fills the gap in UV coverage for
VLBI of the EVN with Hartebeesthoek.  It is 6\arcdeg\ north of the
equator and so it can see the entire plane of the
Milky Way and almost the entire sky.

\subsection{The antenna conversion process}

The drive systems of out of service antennas will generally need some
rectification, and this is in process with the Kuntunse antenna.

More challenging is the development of radio astronomy receiver systems.
The initial aim is to build receivers for operation in radio
astronomy bands within the original Intelsat band range, possibly
re-using the existing microwave feed horn.  The most commonly used frequency
band for VLBI in the EVN is 4.8 -- 5.1 GHz; it falls between the current
transmit and receive bands of the antenna and should not be difficult to
implement with the existing feed.  The 6.7 GHz methanol maser line lies a
little above the feed design upper frequency and tests will be needed to see
if it works at this frequency.  Wide (octave) band feeds and multi-band
feeds are now commonplace.  A feed covering 4.5 to 9 GHz, either
continuously or only in the relevant radio astronomy bands, would include the
two bands already discussed and the 8.4 GHz VLBI band and would permit a wide
range of science to be done.

Various options exist for developing capability outside these bands. The
antenna design is the so-called ``beam waveguide'' in which large diameter
pipes with 45\arcdeg\ mirrors pass the signal to a focus in the room below the
antenna, where the receiver is located.  Most of the African antennas are of
this design.  It comes with some advantages and disadvantages compared to
the Cassegrain design widely used on radio telescopes of this size, but the
technical issues are solvable.

\section{Summary}

The potential for converting obsolete large antennas for radio astronomy has been recognised
World-wide.  The opportunity exists for African countries to re-use these
expensive assets that are becoming redundant
to promote science development on the continent, at relatively low cost.

\end{document}